\documentstyle[preprint,aps]{revtex}
\begin{document}
\draft
\title{Quasiclassical approach to impurity effect on
magnetooscillations in 2D metals.}
\author{A.D. Mirlin$\sp{1,2}$, E. Altshuler$\sp{3}$,
 and  P. W\"{o}lfle$\sp{1}$}
\address{
$\sp{1}$ Institut f\"{u}r Theorie der Kondensierten Materie,
  Universit\"{a}t Karlsruhe, 76128 Karlsruhe, Germany}
\address{
$^2$ Petersburg Nuclear Physics Institute, 188350 Gatchina, St.Petersburg,
Russia.}
\address{ $\sp{3}$ Department of Condensed Matter Physics,
The Weizmann Institute of Science, 76100 Rehovot, Israel
}
\date{\today}
\maketitle
\tighten
\begin{abstract}
We develop a quasiclassical method based on the path integral
formalism, to study the influence of disorder on magnetooscillations
of the density of states and conductivity. The treatment is
appropriate for electron systems in the presence of a random potential
with large
correlation length or a random magnetic field, which are
characterisitic features of various 2D electronic systems presently
studied in experiment. In particular, we study the system of composite
fermions in the fractional quantum Hall effect device, which are
coupled to the Chern--Simons field and subject to a long--range random
potential.
\end{abstract}

\section{Introduction.}

In this article we develop a quasiclassical method  for studying the
influence of disorder on magnetooscillations in 2D electronic
systems. Our original motivation was the problem of quantum particles
in random magnetic fields which has attracted considerable interest
during the last few years.
Only quite recently realizations of a random magnetic
field acting on a 2D electron gas have been prepared and
investigated. One possible way of generating a random magnetic field
is to use a type II disordered superconductor with randomly pinned
flux lines in an external magnetic field as the substrate for the 2D
electron gas \cite{geim}. Alternatively, one may use a magnetically
active substrate such as a demagnetized ferromagnet with randomly
oriented magnetic domains \cite{mancoff}.

The main interest in the random magnetic field problem derives,
however, from effective field theories of  interacting electron
systems, for which the interaction may be shown to be mediated by
fictitious gauge fields. The first example of this class is the gauge
field theory of high--$T_c$ superconductivity compounds
\cite{basand,il,nl,jing,aw2}. There the gauge fields arise as a tool
to implement the projection of the Hilbert space onto a subspace of
states without double occupancy of lattice sites. These gauge fields
are long--range correlated, with correlation function diverging as $1/q^2$
for wavevector $q\to 0$. This corresponds to a short--range correlated
random magnetic field. The second example is provided by the composite
fermion picture of the fractional quantum Hall effect
\cite{jain,lofra,hlr,kz}. There the
electrons are represented by fermions with an even number of magnetic
flux tubes attached. At half-filling of the lowest Landau band, the
average field due to the flux tubes cancels the external
magnetic field, leaving the problem of fermions moving in a random
magnetic field generated by the flux tubes. The correlations of the
corresponding vector potential are also long-ranged.

For a short-range correlated magnetic field, implying a long-range
correlated vector potential, the total scattering rate of a charged
quantum particle diverges in Born approximation \cite{nl,amw}.
This is due to the
divergence of the differential scattering cross-section for forward
scattering. The usual self-consistent treatment of strong scattering
has been shown to be insufficient \cite{1}, as it violates gauge invariance.
The dominance of
small angle scattering calls for a quasiclassical description.
We therefore employ the quasiclassical
approximation method for the path integral representation of this
problem. An additional advantage of the path integral formalism is the
explicit gauge invariance, in contrast to the usual perturbation
theory methods.

For the 2D $GaAs-AlGaAs$ heterostructures on which the fractional quantum
Hall effect (FQHE) \cite{tsui1}
is observed, the quasiclassical treatment is
appropriate for the following resasons.
In these systems, the donors are located
in a remote layer separated by a large distance $d_s\sim 50\div 80 nm$
from the electron gas plane. Thus, the random potential  created
by these impurities
has a large correlation length $\sim d_s$, and the small--angle
scattering dominates, which can be properly described within the
quasiclassical approximation.
 In a strong magnetic field, such that the
Landau level filling factor $\nu$ is close to $1/2$, a statistical
transformation can be applied as mentioned above,
converting the electrons into so-called
composite fermions \cite{jain,lofra,hlr}. Then a random magnetic field
appears, on top of the smooth random potential. The quasiclassical
path integral approach allows to study the effect of both
types of random field on equal footing.

The outline of the paper is as follows. In section 2 we use the path
integral formalism to calculate the total and the transport scattering
rates in long--range correlated random potentials. We check that the
obtained results are in agreement with the conventional perturbation
theory. In section 3 we study the influence of a
long--range random potential
on the oscillations of the density of states and the conductivity in a
magnetic field. In section 4 we generalize the results of the two
preceding sections on the case of a random magnetic field (rather than
random potential). Finally, in section 5 we apply the above formalism
to the system of composite fermions in the FQHE device near $\nu=1/2$,
where both long--range random potential and random magnetic field are
important. Section 6 contains a discussion of the results and
conclusions.

Some of the results of this article have been published in the form of
short communications \cite{1,aamw-fqhe}.

\section{Scattering rates in long--range random potential from the
path integral formalism.}

In this section we develop the quasiclassical path integral formalism
and apply it to the calculation of total and transport scattering
rates in a random potential with large correlation length. These
results can  also be obtained within  conventional perturbation
theory, so that this section has mainly methodological character. It
is instructive, however, to see how the quasiclassical treatment
reproduces results of the perturbation theory, in the case when both
are applicable.

We consider a quantum particle of mass $m$ and energy $E$ moving in 2D
in a static random potential $U(\bbox{r})$ with Gaussian distribution
characterized by the correlation function
\begin{equation}
\langle U(\bbox{r})U(\bbox{r'})\rangle=W(|\bbox{r}-\bbox{r'}|)
\label{1}
\end{equation}
We assume $W(|\bbox{r}|)$ to be a smooth function of $\bbox{r}$
depending on the absolute value $r=|\bbox{r}|$ only, with a
characteristic length scale $\xi\gg\lambda$, where
$\lambda=(2mE)^{-1/2}$ is the wave length (we set $\hbar=1$ throughout
the paper). We review first  results of the standard perturbation
theory. The total scattering rate can be found in the Born
approximation as
\begin{equation}
1/\tau_s=2\pi\int(dp_1)\tilde{W}(|\bbox{p}-\bbox{p_1}|)
\delta(\epsilon_p-\epsilon_{p_1})\ ,
\label{2}
\end{equation}
where $(dp)=d^2p/(2\pi)^2$; $\epsilon_p=p^2/2m$ and
\begin{equation}
\tilde{W}(p)=\int d^2 r\, W(r) \exp(-i\bbox{pr})
\label{3}
\end{equation}
Eq.(\ref{2}) can be transformed as follows
\begin{eqnarray}
1/\tau_s&=&2\pi N(E)\int_0^{2\pi} {d\phi\over 2\pi}
\tilde{W}(2p_F\sin{\phi\over 2})   \nonumber\\
&& = 2\pi N(E) \int_0^{2\pi}d\phi\int_0^\infty dr\,r
J_0\left(2p_Fr\sin{\phi\over 2}\right)W(r)   \nonumber\\
&& = (2\pi)^2 N(E)\int_0^\infty dr\, r J_0^2(p_Fr)W(r)\ ,
\label{4}
\end{eqnarray}
where $ N(E)=m/2\pi$ is the density of states (DOS) on the Fermi surface and
$p_F=(2mE)^{1/2}$ is the Fermi momentum. The above assumption of the
long--range character of $W(r)$, $\xi\gg\lambda$, allows to
approximate the Bessel function in (\ref{4}) by its asymptotic
expression, yielding
\begin{equation}
1/\tau_s= {2\over v_F}\int_0^\infty dr W(r)\ ,
\label{5}
\end{equation}
with $v_F=p_F/m$ being the Fermi velocity. This result holds under the
condition of applicability of the Born approximation, which is
\begin{equation}
v_F\tau_s\gg\xi
\label{5a}
\end{equation}

Expression for the transport scattering rate $1/\tau_{tr}$ differs from
that for $1/\tau_s$ by an additional factor $(1-\cos\phi)$. We have
thus, in full analogy with eq.(\ref{4}),
\begin{eqnarray}
1/\tau_{tr}&=&2\pi N(E)\int_0^{2\pi} {d\phi\over 2\pi}
\tilde{W}(2p_F\sin{\phi\over 2})(1-\cos\phi)   \nonumber\\
&& = 2\pi N(E) \int_0^{2\pi}d\phi\int_0^\infty dr\,r
J_0\left(2p_Fr\sin{\phi\over 2}\right)(1-\cos\phi)W(r)   \nonumber\\
&& = (2\pi)^2 N(E)\int_0^\infty dr\, r [J_0^2(p_Fr)-J_1^2(p_Fr)]W(r)\ .
\label{6}
\end{eqnarray}
Using now the relations $J'_0(z)=-J_1(z)$; $[zJ_1(z)]'=zJ_0(z)$ and
integrating by part, we reduce eq.(\ref{6}) to the form
\begin{equation}
1/\tau_{tr}=-{2\pi^2\over p_F^2} N(E)\int dr J_0^2(p_Fr){d\over
dr}r{d\over dr}W(r).
\label{7}
\end{equation}
For a long range potential, $\xi\gg\lambda$, we find thus
\begin{eqnarray}
1/\tau_{tr}&=&-{m\over p_F^3}\int_0^\infty dr\left[W''(r)+
{W'(r)\over r}\right]
\nonumber \\
&&=-{m\over p_F^3}\int_0^\infty dr {W'(r)\over r}
\label{8}
\end{eqnarray}
In the last line we used the assumption that $W(r)$ is an analytic
function of $\bbox{r}$ at $\bbox{r}=0$, so that $W'(0)=0$.

In the remaining part of this section we rederive eqs.(\ref{5}),
(\ref{8}) in the path integral approach \cite{feinm}.
To calculate $\tau_s$, we
consider the (retarded) single--particle Green function
\begin{equation}
G_R(0,R;T)=\int_{\bbox{r}(0)=0}^{\bbox{r}(T)=\bbox{R}} D\bbox{r}(t)
\exp\left\{i\int_0^Tdt\left[{m\dot{\bbox{r}}^2\over 2} -
U(\bbox{r})\right]\right\}
\label{9}
\end{equation}
After averaging over the disorder (which we will denote by angular
brackets), we get
\begin{equation}
\langle G_R(0,R;T)\rangle=
\int_{\bbox{r}(0)=0}^{\bbox{r}(T)=\bbox{R}} D\bbox{r}(t)
\exp\left\{i\int_0^Tdt{m\dot{\bbox{r}}^2\over 2} - {1\over 2}
\int_0^T \int_0^T dt\, dt'\,W[\bbox{r}(t)-\bbox{r}(t')]\right\}
\label{10}
\end{equation}
In the absence of the second term in the exponent, eq.(\ref{10})
describes the free particle Green function
$\langle G^{(0)}_R(0,R;T)\rangle$, which can be found exactly by the
saddle--point method. The saddle point is given by the classical
trajectory $\bbox{r}(t)=\bbox{v}t$; $\bbox{v}=\bbox{R}/T$, which
yields in 2D
\begin{equation}
\langle G^{(0)}_R(0,R;T)\rangle={m\over 2\pi i T}e^{imR^2/2T}
\label{11}
\end{equation}
We use now the fact that the correlator $W(r)$ is smooth, and thus the
full expression (\ref{10}) can still be evaluated
quaiclassically. Furthermore, we assume the random potential to be
relatively weak, so that the second term in the action in (\ref{10})
can be considered as a perturbation. Formally, this means $E\tau_s\gg
1$. Then the effect of the random potential term is given by its value
on the saddle--point trajectory. Assuming now $v_F\tau_s\gg\xi$, as
in eq.(\ref{6}), we get
\begin{eqnarray}
\langle G_R(R;T)\rangle&=&\langle G^{(0)}_R(0,R;T)\rangle
\exp\left(-T\int_{0}^{\infty}{dr\over v_F} W(r)\right)
\nonumber\\
&&=\langle G^{(0)}_R(0,R;T)\rangle \exp(-T/2\tau_s)\ ,
\label{12}
\end{eqnarray}
with $\tau_s$ as in eq.(\ref{5}).
We find therefore that {\it the single particle relaxation rate can be
simply found by evaluating the random potential--induced term in the
action on a classical trajectory.}

As we will see now, calculation of the transport time in this
formalism is much more elaborate. We start from the Kubo formula for
the conductivity
\begin{equation}
\sigma=-{e^2\over 4\pi} {1\over V}\langle\mbox{Tr}
\hat{v}_x[G_R(E)-G_A(E)]\hat{v}_x[G_R(E)-G_A(E)]\rangle
\label{13}
\end{equation}
where $V$ is the system volume, $\hat{v}_x$ is velocity operator and
$G_R$, $G_A$ denote retarded and advanced Green functions respectively.
As usual, the leading contribution is given by the terms $\sim G_R
G_A$, which can be rewritten in time representation as
\begin{equation}
\sigma={e^2\over2\pi}\int_0^\infty dT_1 \int_0^\infty dT_2 \int d^2R
\langle\hat{v}_xG_R(0,R;T_1)\hat{v}_xG_A(R,0;-T_2)\rangle
e^{iE(T_1-T_2)}
\label{14}
\end{equation}
The product of the  Green function can be expressed in terms of the
path integral, in analogy with eq.(\ref{10}) (we omit here the vertex
velocity operators for simplicity and restore them in the end of
calculation):
\begin{eqnarray}
&&\langle G_R(0,R;T_1)G_A(R,0;-T_2)\rangle \nonumber\\ && =
\int_{\bbox{r_1}(0)=0}^{\bbox{r_1}(T_1)=\bbox{R}} D\bbox{r_1}(t)
\int_{\bbox{r_2}(0)=0}^{\bbox{r_2}(T_2)=\bbox{R}} D\bbox{r_2}(t)
\exp\left\{i\int_0^{T_1}dt {m\dot{\bbox{r_1}}^2\over 2}
-i\int_0^{T_2}dt {m\dot{\bbox{r_2}}^2\over 2} \right. \nonumber\\
 &&
- {1\over 2} \int_0^{T_1} \int_0^{T_1} dt\, dt'\,
W[\bbox{r_1}(t)-\bbox{r_1}(t')]
- {1\over 2} \int_0^{T_2} \int_0^{T_2} dt\, dt'\,
W[\bbox{r_2}(t)-\bbox{r_2}(t')] \nonumber \\ && \left.
+ \int_0^{T_1} \int_0^{T_2} dt\, dt'\,
W[\bbox{r_1}(t)-\bbox{r_2}(t')]
\right\}
\label{15}
\end{eqnarray}
One may attempt to evaluate eq.(\ref{15}) by the same simple
saddle--point method, which led us from eq.(\ref{10}) to
eq.(\ref{11}). The saddle point trajectory for the action in
(\ref{15}) is
\begin{equation}
\bbox{r_1}^c(t)={t\over T_1}\bbox{R}\ ;\qquad
\bbox{r_2}^c(t)={t\over T_2}\bbox{R}
\label{16}
\end{equation}
However, this does not give the leading contribution to
(\ref{15}). In mathematical terms, the failure of the usual
 saddle point approximation  is related to the fact that
for $T_1$ and $T_2$
close to each other (this condition is implied by the factor
$\exp\{iE(T_1-T_2)\}$ in eq.(\ref{14}))
the impurity--induced terms in the action
in (\ref{15}) nearly cancel each other on this trajectory.
Moreover, this happens always, when $\bbox{r_1}$ and $\bbox{r_2}$
traverse  the same trajectory in the same time, so that the vicinities of
all such trajectories may be expected to be relevant. Physically speaking,
this is because the conductivity is determined by the time scale in which
the direction of velocity of the particle changes by an angle of order
$\pi$. Since in a long--range potential  small--angle scattering is
typical,  this happens after many scattering events. The corresponding
classical trajectory is smooth, but globally is typically very different from a
straight line. These considerations suggest to study contributions
from a class of paths  $\bbox{r_1}(t)$ and
$\bbox{r_2}(t)$, which fluctuate by a small amount about a common smooth,
but otherwise arbitrary, trajectory $\bbox{\rho}(t)$,
 traversed with a constant (by absolute value) velocity
$v_F$. Though these
trajectories are not really saddle points of the action, they are
``nearly saddle points'' and can be expected to dominate the path
integral. Thus, we represent the two paths  $\bbox{r_1}(t)$ and
$\bbox{r_2}(t)$ in (\ref{15}) in the form
\begin{eqnarray}
&&\bbox{r_1}(t_1)=\bbox{\rho}(t)+{1\over 2}\bbox{r_-}(t)\ ;
\nonumber\\
&&\bbox{r_2}(t_2)=\bbox{\rho}(t)-{1\over 2}\bbox{r_-}(t)\ ;
\nonumber\\
&& t_1= t{T_1\over t_+}=t{t_+ + t_-/2\over t_+ }\ ;\qquad
t_2=t{T_2\over t_+}=t{t_+ - t_-/2\over t_+ }\ ,
\label{16a}
\end{eqnarray}
where we define $t_+=(T_1+T_2)/2$ and $t_-=T_1-T_2$.
The variable $t$ in eq.(\ref{16a}) is confined to the interval $[0,t_+]$.
 Note that
the typical values of $t_-$ and $t_+$
are expected to be $t_-\sim 1/E$ and $t_+\sim
\tau_{tr}$, so that $t_-\ll t_+$.
Now we expand the action (\ref{15}) in $\bbox{r}_-(t)$. The leading
contribution from the kinetic energy terms reads:
\begin{eqnarray}
S_{kin}&=&-i\int_0^{t_+ + t_-/2}dt_1 {m\over 2}\bbox{\dot{r}}_1^2+
        i\int_0^{t_+ - t_-/2}dt_2 {m\over 2}\dot{\bbox{r}}_2^2
       \nonumber \\
       &=&-{im\over 2}{t_-\over t_+}\int_0^{t_+}dt\dot{\bbox{\rho}}^2 +
\delta S_{kin} \ ;  \nonumber \\
\delta S_{kin} &\simeq& -im\int_0^{t_+}dt\dot{\bbox{\rho}}(t)
\bbox{\dot{r}_-}(t)
\label{17}
\end{eqnarray}
For the disorder--induced terms in the action (\ref{15}) we use the
Taylor expansion
\begin{eqnarray}
&& W(|\bbox{r}+\bbox{\delta r}|)\simeq W(r)+\partial_iW(r)\delta r_i
+{1\over 2}\partial_i\partial_jW(r)\delta r_i\delta r_j\ ; \nonumber\\
&&\partial_i W(r)=W'(r){r_i\over r}\ ; \nonumber\\
&&\partial_i\partial_j W(r)={W'(r)\over r}
\left(\delta_{ij}-{r_i r_j\over r^2}\right)+ W''(r){r_i r_j\over r^2}\ ,
\label{18}
\end{eqnarray}
with $\bbox{r}=\bbox{\rho}(t)-\bbox{\rho}(t')$ and
$\bbox{\delta r}={1\over 2}[\pm\bbox{r_-}(t)\pm\bbox{r_-}(t')]$.
Then the last three terms in (\ref{15}) combine to
\begin{eqnarray}
\delta S_W & \simeq & -{1\over 2}\int_0^{t_+}\int_0^{t_+}dt\,dt'
{W'[|\bbox{\rho}(t)-\bbox{\rho}(t')|]\over|\bbox{\rho}(t)-\bbox{\rho}(t')|}
r_-^\bot(t)r_-^\bot(t')
\nonumber \\
&-&{1\over 2}\int_0^{t_+}\int_0^{t_+}dt\,dt'
W''[|\bbox{\rho}(t)-\bbox{\rho}(t')|]
r_-^\|(t)r_-^\|(t')
\label{19}
\end{eqnarray}
where we separated the fluctuations $\bbox{r}_-(t)$ into the longitudinal
$r_-^\|(t)$ and transverse $r_-^\bot(t)$ parts with respect to the
direction of velocity $\dot{\bbox{\rho}}(t)$. As is expected on
physical grounds and will be justified below, the relevant
trajectories have  nearly constant absolute value of velocity
$|\dot{\bbox{\rho}}(t)|=v$. We also expect, in view of eq.(\ref{5a}),
the fluctuations $\bbox{r_-}(t)$ to be varying slowly on
the time scale $\xi/v$.
Then eq.(\ref{19}) reduces to
\begin{equation}
\delta S_w\simeq {1\over v} w_0^\bot\int_0^{t^+}dt
r_-^\bot(t)r_-^\bot(t)\ ,
\label{20}
\end{equation}
with $w_0^\bot=-\int_0^\infty {W'(r)\over r}dr$ and assuming again
$\int W''dr=0$. Finally, substituting
(\ref{17}), (\ref{20}) into eq.(\ref{14}), we get
\begin{eqnarray}
\sigma&=&{e^2\over 4\pi}\int_0^\infty dt_+ \int_{-\infty}^\infty dt_-
\int d^2R\, e^{iEt_-} \int D\bbox{\rho}(t)\int Dr_-^\bot(t)\int Dr_-^\|(t)
\dot{\bbox{\rho}}(0)\dot{\bbox{\rho}}(t_+)
\nonumber\\
&\times& \exp\left\{-{im\over 2}{t_-\over t_+}\int_0^{t_+}dt
\dot{\bbox{\rho}}^2  -im\int_0^{t_+}dt\dot{\bbox{\rho}}(t)
\bbox{\dot{r}_-}(t)-{1\over v} w_0^\bot\int_0^{t^+}dt
r_-^\bot(t)r_-^\bot(t)\right\}
\label{21}
\end{eqnarray}
We integrate first over the longitudinal fluctuations
$r_-^\|(t)$. This produces the $\delta$--function
$\prod_t\delta(\ddot{\bbox{\rho}}(t)^\|)$, thus restricting the integral to
the trajectories $\rho(t)$ with constant velocity
$v=|\dot{\bbox{\rho}}(t)|$, as was expected. Further, the integral
over $t_-$ gives then $\delta(E-mv^2/2)$, fixing the value of $v$.
Now we take the integral over $r_-^\bot(t)$:
\begin{eqnarray}
&&\int Dr_-(t)\exp\{im\int_0^{t_+}dt[\dot{\bbox{v}}(t)]^\bot
r_-^\bot(t)-{1\over v} w_0^\bot\int_0^{t^+}dt
r_-^\bot(t)r_-^\bot(t)\}
\nonumber\\
\qquad &\sim& \exp\left\{-{m^2v_F\over 4w_0^\bot}\int dt
\dot{\bbox{v}}^2(t)\right\} \nonumber \\
&=& \exp\left\{-{m^2v_F^3\over 4w_0^\bot}\int dt
\dot{\phi}^2(t)\right\}\ ,
\label{22}
\end{eqnarray}
where $\phi$ is the polar angle of the velocity vector $\bbox{v}$.
The action (\ref{22}) corresponds to the Fokker--Planck (diffusion)
process for the angle $\phi$:
\begin{equation}
\langle\dot{\phi}(t)\dot{\phi}(t')\rangle={2w_0^\bot\over m^2v_F^3}
\delta(t-t')\ ,
\label{23}
\end{equation}
or, in discrete version,
\begin{equation}
\langle\delta\phi^2\rangle={2w_0^\bot\over m^2v_F^3}\delta t
\label{24}
\end{equation}
The corresponding Fokker--Planck equation for the distribution
function $P(\phi,t)$ reads
\begin{equation}
{w_0^\bot\over m^2v_F^3} {\partial^2\over\partial\phi^2} P(\phi,t)=
{\partial\over\partial t}P(\phi,t)
\label{25}
\end{equation}
Thus,
\begin{equation}
\sigma=e^2 N(E){v_F^2\over 2}\int dt_+\int d\phi \cos\phi P(\phi,t_+)\ ,
\label{26}
\end{equation}
where $P(\phi,t)$ satisfies eq.(\ref{25}) and the boundary condition
$P(\phi,0)=\delta(\phi)$. To fix the normalization in eq.(\ref{26}), we have
exploited the condition of the particle number conservation:
\begin{equation}
\int d^2R\int_{-\infty}^{\infty}dt_-e^{iEt_-}
\langle G_R(0,R;T_1)G_A(R,0;-T_2)\rangle =2\pi
N(E).
\label{26a}
\end{equation}
The solution of eq.(\ref{25}) has the form
\begin{equation}
P(\phi,t)=\sum_{m=-\infty}^{\infty}\exp\left\{im\phi-{m^2v_F^3\over
w_0^\bot}t\right\}.
\label{26b}
\end{equation}
We get therefore $\sigma=e^2 N(E) v_F^2\tau_{tr}/2$,
with
\begin{equation}
{1\over\tau_{tr}}={w_0^\bot\over m^2v_F^3}\ ,
\label{27}
\end{equation}
in precise agreement with eq.(\ref{8})

We have seen therefore how the quasiclassical treatment of impurity
potential reproduces the results of perturbation theory. In the next
section we apply the method to the situation when the perturbation
theory breaks down.

\section{Magnetooscillations in the presence of long--range random
potential.}

We consider a 2D gas of charged particles subject to a uniform
magnetic field $B$ and a smooth random potential $U(\bbox{r})$ defined
by eq.(\ref{1}). We will assume the impurity scattering to be
relatively weak, so that $\omega_c\tau_{tr}\gg 1$, where
$\omega_c=eB/mc$ is the cyclotron frequency. Our quasiclassical
treatment will be valid for a random potential with correlation length
\begin{equation}
\xi\gg l_B\ ,
\label{28}
\end{equation}
 where $l_B=(c/eB)^{1/2}$ is the magnetic length. Eq.(\ref{28}) is
opposite to the condition of applicability of the self-consistent Born
approximation \cite{ando,raikh,la}. The de Haas--van Alphen oscillations
(dHvAO)  of the density of states (DOS) were studied in this regime
in \cite{raikh} by approximate summation of all orders of perturbation
theory. We will demonstrate
 that this can be achieved in a more elegant way
from the path integral formalism. We will also show that the
Shubnikov--de Haas oscillations (SdHO) of conductivity can be
described in this way, as well.

We consider first the DOS, which can be found from the
single--particle Green function (\ref{10}) as
\begin{eqnarray}
&&\rho(E)=-{1\over\pi}\mbox{Im}G_R(E)\ ; \nonumber \\
&&G_R(E)=\int_0^\infty\langle G_R(0,0;T)\rangle e^{iET}dT
\label{29}
\end{eqnarray}
In the quasiclassical approximation, the Green function $G_R(E)$ can
be represented as a sum over closed classical orbits \cite{reichl}
\begin{equation}
G_R(E)={m\over 2}
\left\{ {1\over\pi}\ln (-E) - i\sum_{\beta}
D_{\beta}\exp iS_{\beta}(E)\right\} ,
\label{30}
\end{equation}
where $\beta$ labels the orbits, $D\sb{\beta}$
is a factor originating from the path integration over
the vicinity of the classical orbit $\beta$, and
$S\sb{\beta}(E)$ is an action in the energy representation.
In the absence of the random potential, we would
have just a free particle in uniform magnetic field. The classical
trajectories are then the cyclotron orbits with radius
$R_c=v_F/\omega_c$. They can be classified by the winding number $k$,
specifying the number of times the orbit is traversed. Eq.(\ref{30})
takes then the form
\begin{equation}
G_R^{(0)}(E)={m\over 2 }
\left[{1\over\pi}\ln(-E)-2i\sum_{k=1}^{\infty}
\exp { \left\{ 2\pi k i\left[ {E\over \omega_c} +{1\over 2}\right]
\right\} }\theta(E)\right]\ ,
\label{31}
\end{equation}
where $\theta(E)$ is the step function.
In particular, it is easy to check by using the Poisson resummation
formula that (\ref{31}) gives the correct expression for the DOS in
terms of the sum over Landau levels:
\begin{equation}
\rho(E)={1\over 2\pi l_B^ 2}\sum_{N=0}^{\infty}
\delta[E- \omega_c(N +{1/2})]
\label{32}
\end{equation}
Since the impurity scattering is assumed to be weak, we neglect its
influence on the classical trajectories and on the prefactor $D_\beta$,
in full analogy with calculation of $\tau_s$ in Section 2. This gives,
instead of (\ref{31}),
\begin{equation}
G_R(E)={m\over 2 }
\left[{1\over\pi}\ln(-E)-2i\sum_{k=1}^{\infty}
\exp  \left\{ 2\pi k i\left[ {E\over \omega_c} +{1\over 2}\right]
-S_W k^2\right\} \theta(E)\right]\ ,
\label{33}
\end{equation}
where $S_W$ is given by the second term in the action in (\ref{10})
evaluated on a cyclotron orbit of winding number $k=1$. It is easily
found to be equal to
\begin{eqnarray}
S_W&=&{1\over 2v_F^2}\oint dr\oint dr' W(|\bbox{r}-\bbox{r'}|)
\nonumber\\
&=&{1\over 2v_F^2}\int(dq)\tilde{W}(q)[2\pi R_c J_0(qR_c)]^2
\nonumber\\
&=&{\pi\over \omega_c^2}\int dq\, q\tilde{W}(q)J_0^2(qR_c)
\label{34}
\end{eqnarray}
If the correlation length $\xi$ of the random potential satisfies the
condition $\xi\ll R_c$, $S_W$ takes the form
\begin{equation}
S_W={1\over\omega_c v_F}\int_0^\infty dq \tilde{W}(q)={2\pi\over
v_F\omega_c} \int_0^\infty dr W(r)={\pi\over\omega_c\tau_s}\ ,
\label{35}
\end{equation}
with $\tau_s$ as found in section 1, see eq.(\ref{5}). We get then for
the DOS at $E>0$
\begin{equation}
\rho(E)={m\over 2\pi}\left[1+2\sum_{k=1}^\infty(-1)^k\cos\left(2\pi
k{E\over\omega_c}
\right)\exp\left(-k^2{\pi\over\omega_c\tau_s}\right)\right]\ ,
\label{36}
\end{equation}
or after resumming by the Poisson formula,
\begin{equation}
\rho(E)={1\over 2\pi l_B^ 2}\sum_{N=0}^{\infty}
\sqrt{{\tau_s\over\omega_c}}\exp\left\{-{\pi\tau_s\over\omega_c}
[E- \omega_c(N +{1/2})]^2\right\}
\label{37}
\end{equation}
This formula shows that the Landau levels acquire  a Gaussian
form. They are well resolved if $\omega_c\tau_s\gg 1$. In the opposite
case, $\omega_c\tau_s\ll 1$, the representation (\ref{36}) is
appropriate, where all harmonics except the first one can be omitted:
\begin{eqnarray}
\rho(E)&=&{m/2\pi}+\rho_{osc}(E)\ ;\nonumber\\
\rho_{osc}&\simeq&-{m\over\pi}\cos(2\pi E/\omega_c)e^{-\pi/\omega_c\tau_s}
\label{38}
\end{eqnarray}
The dependence of the amplitude of oscillations on the magnetic field
has the same form $\sim\exp(-\pi/\omega_c\tau_s)$  as for the short
range potential, so that if one plots $\log\rho_{osc}$ versus $1/B$
(so-called Dingle plot \cite{shoenberg}), one expects to get a linear behavior.

In the case of ultra-long-range potential with $\xi\gg R_c$, we find
$S_W=2\pi^2W(0)/\omega_c^2$. The Landau levels have again a Gaussian
shape:
\begin{equation}
\rho(E)={1\over 2\pi l_B^ 2}\sum_{N=0}^{\infty}
\sqrt{2\pi W(0)}\exp\left\{-{1\over 2W(0)}
[E- \omega_c(N +{1/2})]^2\right\}\ ,
\label{39}
\end{equation}
 which in this case reflects their inhomogeneous broadening. The
oscillating part of the DOS in eq.(\ref{38}), $\rho_{osc}$, is now equal
to
\begin{equation}
\rho_{osc} \simeq -{m\over\pi}\cos(2\pi E/\omega_c)
e^{-2\pi^2W(0)/\omega_c^2}\ ,
\label{40}
\end{equation}
so that the Dingle plot is expected to be quadratic.

As we have already mentioned, the above results for the DOS
oscillations in a long range potential were obtained in \cite{raikh}
by resummation of the perturbation theory expansion \cite{Gerh}.
Besides being
simpler and physically more transparent, the present derivation has the
advantage that it can be straightforwardly generalized to SdHO of
conductivity or to the case of random magnetic field.

The conductivity oscillations were recently considered in the
framework of the path integral approach for the case of short--range
random potential ($\tau_{tr}=\tau_s$) in \cite{opp,rich}. The authors
of these papers started from the Kubo formula (\ref{14}), representing
each Green function as a sum over classical trajectories in uniform
magnetic field and taking scattering into account by including the
factor $\exp(-t/\tau_s)$. A trajectory is characterized by a number of
cyclotron revolutions $k$. Therefore, eq.(\ref{14}) is reduced to a
double sum over the winding numbers $k_R,\ k_A$. The non-oscillating
contribution to the conductivity corresponds then to the terms with
$k_R=k_A$, whereas the $n$-th harmonic of the oscillations is
described by the terms with $|k_R-k_A|=n$. This holds for our problem
of smooth random potential as well. The only difference is that the
relevant trajectories are now not exactly cyclotron circles, but
rather are very close to them, however drifting a little from one
revolution to another. The typical total number of revolutions of each
trajectory is of order of $\omega_c\tau_{tr}\gg 1$; in this time the shift
of the center of cyclotron movement for a typical trajectory is of
order $R_c$. This is completely analogous to the consideration of
conductivity in zero magnetic field in section 2, where the
characteristic trajectories dominating the path integral were smooth
(i.e. {\it locally} close to straight lines), but not really straight
lines.

The non-oscillating contribution $\sigma_0$ to the conductivity
is given by the trajectories for $G_R$ and $G_A$ having equal number of
cyclotron revolutions and following closely each other within a
distance $\ll\xi$, as in section 2. Then the impurity--induced terms
in (\ref{15})  cancel each other again to a great extent, leading to
$\tau_{tr}\gg\tau_s$. The exact evaluation of the conductivity $\sigma_0$
is however much more complicated in the present situation, since the
trajectories return to positions close to the preceding ones
after $1,2,\ldots$ revolutions
and may interact through the impurity correlator. This may lead to
a deviation of $\tau_{tr}$ from its value calculated in zero magnetic
field in section 2. We leave this problem aside in the present article and
concentrate on the exponential damping of the oscillations by disorder.

The leading contribution to oscillations (the first harmonic)
corresponds to the case when one of the trajectories has one extra
cyclotron revolution as compared to another one. Then the
contribution to the effective action in (\ref{15}) from this part of the
trajectory is not cancelled and leads to a suppression of the
amplitude of oscillations. We get
\begin{equation}
\sigma=\sigma_0+\sigma_1\cos(2\pi E/\omega_c)e^{-S_W}+\ldots\ ,
\label{41}
\end{equation}
with $S_W$ given by eq.(\ref{34}). Thus, impurity scattering leads to
the same $e^{-S_W}$ exponential damping of the SdHO, as for the DOS
oscillations. Depending on the relation between $\xi$ and $R_c$, this
factor takes the form
\begin{equation}
e^{-S_W}=\left\{
\begin{array}{ll}
e^{-\pi/\omega_c\tau_s}\ ,\qquad & \xi\ll R_c\\
e^{-2\pi^2W(0)/\omega_c^2}\ ,\qquad & \xi\gg R_c
\end{array}
\right.
\label{42}
\end{equation}
as in eqs. (\ref{38}), (\ref{40}).

\section{Conductivity and magnetooscillations in random magnetic field.}

In this section we consider the oscillations of DOS and conductivity
in the case when the carriers are subject to a random magnetic field
rather than to a random potential.
The properties of a quantum particle moving in a random magnetic field
have been intensively discussed in the recent literature
\cite{amw,ai,km,Pry,Sug,Avi,KWAZ,ZA,chalk,gav,bar,ublee}.
We are aware by now of three completely different physical systems,
for which the random magnetic
field problem is relevant. In the first class of system
the random magnetic field is generated by a substrate on top of which
the 2D electron gas is placed. Realizations of the substrate are a
type II superconductor with randomly pinned flux lines \cite{geim}, or
a demagnetized ferromagnet with randomly oriented domains
\cite{mancoff}. The second is the state with
spin--charge separation of high--$T_c$ superconducting materials
\cite{basand,il,nl,jing}. To describe it, one introduces a fictitious
$U(1)$ gauge field interacting with charge carriers. Here, the transverse
(magnetic) component of the gauge field is the most important.
The existence of gauge field fluctuations in
high-$T\sb{c}$ compounds can be inferred \cite{aw2} from experimental
observations of unusual weak localization corrections to
the magnetoconductance \cite{jing}. Finally, the
third application concerns the FQHE system in the vicinity of
$\nu=1/2$ filling factor of the Landau level. This will be studied
in more detail in the next section.

We consider a Gaussian distributed random magnetic field $h(\bbox{r})$
(perpendicular to the 2D plane) with the correlator
\begin{equation}
\left({e\over c}\right)^2\langle h(\bbox{r}) h(\bbox{r'})\rangle =
\Gamma\delta^{(2)}(\bbox{r}-\bbox{r'})
\label{43}
\end{equation}
We will assume the weak disorder case, which means $\Gamma\ll mE_F$.
Let us first reconsider the evaluation of scattering rates in section
2 for this type of disorder. For the one-particle Green function we
have, instead of (\ref{10}):
\begin{eqnarray}
\langle G_R(0,R;T)\rangle&=&
\int_{\bbox{r}(0)=0}^{\bbox{r}(T)=\bbox{R}} D\bbox{r}(t)
\exp\left\{i\int_0^Tdt{m\dot{\bbox{r}}^2\over 2}\right. \nonumber\\ &-&
\left.{1\over 2}\left({e\over c}\right)^2
\int_0^T \int_0^T dt\, dt'\, \dot{r}_i(t)\dot{r}_j(t')
\langle a_i(\bbox{r}(t))a_j(\bbox{r}(t'))\rangle\right\}\ ,
\label{44}
\end{eqnarray}
 where $\bbox{a}(\bbox{r})$ is a vector potential corresponding to the
magnetic field $h(\bbox{r})$. However, the correlator $\langle
a_i a_j\rangle$ is gauge-dependent, so that eq.(\ref{44})
does not allow to define meaningfully the single particle relaxation
time $\tau_s$, as it was possible for the random potential case.
We will discuss the problem of definiton of $\tau_s$ below.

For the transport time, the problem of gauge invariance does not
apply. The analogue of eq. (\ref{15}) reads
\begin{eqnarray}
&& \langle G_R(0,R;T_1)G_A(R,0;-T_2)\rangle \nonumber\\ && =
\int_{\bbox{r_1}(0)=0}^{\bbox{r_1}(T_1)=\bbox{R}} D\bbox{r_1}(t)
\int_{\bbox{r_2}(0)=0}^{\bbox{r_2}(T_2)=\bbox{R}} D\bbox{r_2}(t)
\exp\left\{i\int_0^{T_1}dt {m\dot{\bbox{r_1}}^2\over 2}
-i\int_0^{T_2}dt {m\dot{\bbox{r}}^2\over 2} \right. \nonumber\\
 &&
- {1\over 2}\left({e\over c}\right)^2 \int_0^{T_1} \int_0^{T_1} dt\, dt'\,
\dot{r}_{1i}(t)\dot{r}_{1j}(t')
\langle a_i(\bbox{r_1}(t))a_j(\bbox{r_1}(t'))\rangle\nonumber\\
&& - {1\over 2} \left({e\over c}\right)^2\int_0^{T_2} \int_0^{T_2} dt\, dt'\,
\dot{r}_{2i}(t)\dot{r}_{2j}(t')
\langle a_i(\bbox{r_2}(t))a_j(\bbox{r_2}(t'))\rangle
\nonumber \\ && \left.
+ \left({e\over c}\right)^2\int_0^{T_1} \int_0^{T_2} dt\, dt'\,
\dot{r}_{1i}(t)\dot{r}_{2j}(t')
\langle a_i(\bbox{r_1}(t))a_j(\bbox{r_2}(t'))\rangle
\right\} \nonumber \\
&& =
\int_{\bbox{r_1}(0)=0}^{\bbox{r_1}(T_1)=\bbox{R}} D\bbox{r_1}(t)
\int_{\bbox{r_2}(0)=0}^{\bbox{r_2}(T_2)=\bbox{R}} D\bbox{r_2}(t)
\exp\left\{i\int_0^{T_1}dt {m\dot{\bbox{r_1}}^2\over 2}
-i\int_0^{T_2}dt {m\dot{\bbox{r}}^2\over 2} -{1\over 2}\Gamma
s_{no}\right\}
\label{45}
\end{eqnarray}
where we used the Stokes theorem to rewrite the effective action in
the explicitly gauge-invariant form. Here $s_{no}$ is so-called
non-oriented, or Amp\`erean, area inside the closed curve formed by
the two trajectories $\bbox{r_1}(t)$ and $\bbox{r_2}(t)$ \cite{wheat}.
We recall, in particular, that the non-oriented area enclosed by $k$
windings of
a trajectory is the geometrical area multiplied by $k^2$.
We repeat now the derivation of $\tau_{tr}$ performed in section 2. For
the disorder-induced part of the action we find, instead of
(\ref{20}),
\begin{equation}
\delta S_W= {v\Gamma\over 2}\int_0^{t_+}dt|r_-^\bot(t)|
\label{46}
\end{equation}
Integral over the fluctuations $r_-(t)$, eq.(\ref{22}), now has the
following form:
\begin{eqnarray}
&&\int Dr_-^\bot(t)\exp\{-im\int dt [\dot{\bbox{v}}]^\bot r_-^\bot -
{\Gamma v\over 2}\int dt |r_-|\} \nonumber \\
&&\qquad = \prod_{\delta t_i}\int dr_-^{(i)}\exp\left\{\delta
t_i\left(-im[\dot{\bbox{v}}(t_i)]^\bot r_-^{\bot(i)}
-{\Gamma v\over 2}|r_-^{\bot(i)}|\right)\right\}
\nonumber \\
&&\qquad \prod_{\delta t_i} \frac
{\Gamma v\,\delta t_i} {[(\Gamma v/2)\delta t_i]^2 + m^2[\delta
\bbox{v}(t_i)]^2}
\label{47}
\end{eqnarray}
This implies that the scattering angle
$\delta\phi\simeq[\delta \bbox{v}]^\bot/v$
obeys the Cauchy distribution
\begin{equation}
P(\delta\phi)={1\over\pi} \frac {(\Gamma/2m)\delta t}
{(\Gamma/2m)^2(\delta t)^2+(\delta\phi)^2}
\label{48}
\end{equation}
The Boltzmann equation corresponding to eq.(\ref{48}) is found to be
\begin{equation}
{\partial P(\phi,t)\over \partial t} = \int d\phi' w(\phi-\phi')
[P(\phi',t)-P(\phi,t)]\ ,
\label{49}
\end{equation}
with
\begin{equation}
w(\phi)={1\over\pi} {\Gamma\over 2m} {1\over \phi^2}\ ,\qquad
\phi\ll\pi
\label{50}
\end{equation}
This can be easily checked by solving eq.(\ref{49}) by means of the
Fourier transform in the $\phi$--space. The quasiclassical method
describes correctly only the small angle scattering, so it is able to
give an expression for the differential scattering rate $w(\phi)$ for
$\phi\ll\pi$ only. The transport scattering rate
$1/\tau_{tr}=\int_{-\pi}^\pi w(\phi)(1-\cos\phi)d\phi$ can be found in
this way up to a numerical coefficient only: $1/\tau_{tr}\sim
\Gamma/m$. The exact value can be found from the perturbation theory
calculation, which gives \cite{amw}
\begin{equation}
w(\phi)={\Gamma\over 8\pi m} \cot^2\phi/2\ ,
\label{51}
\end{equation}
in full agreement with eq.(\ref{50}). Consequently,
\begin{equation}
1/\tau_{tr}=\Gamma/4m
\label{52}
\end{equation}

As we can see from eqs.(\ref{50}), (\ref{51}), the total scattering
rate $1/\tau_s=\int d\phi w(\phi)$ diverges at $\phi\to 0$. This is
related to the fact that the contribution $S_h$ to the action from the
random magnetic field is proportional to the area, rather than to the
length, of the trajectory. We will see below, when studying
 dHvAO and SdHO, that the cyclotron motion of the particle provides a
natural regularization of this divergency. However, this
regularization is determined by the geometry of the experiment
considered. Thus, {\it the single particle relaxation time in the
random magnetic field is dependent on the geometry of the problem}.

We turn now to the consideration of magnetooscillations in the case when a
uniform magnetic field $B$ is applied in addition to the random
one. As in  section 3, we will assume a relatively strong field,
 meaning $\omega_c\tau_{tr}\gg 1$, or according to eq.(\ref{52}),
$m\omega_c\gg \Gamma$. In full analogy with eq.(\ref{36}), the DOS can
then be written as
\begin{equation}
\rho(E)={m\over 2\pi}\left[1+2\sum_{k=1}^\infty(-1)^k\cos\left(2\pi
k{E\over\omega_c}
\right)\exp\left(-k^2 S_h\right)\right]\ ,
\label{53}
\end{equation}
with $S_h$ being the random magnetic field action on a simple (winding
number 1) cyclotron orbit. For a $\delta$--like correlated magnetic
field, eq.(\ref{43}), it is equal to
\begin{equation}
S_h={1\over 2}\Gamma s_{no}={1\over 2}\Gamma\pi R_c^2=
\pi E\Gamma/m\omega_c^2
\label{54}
\end{equation}
Resumming eq.(\ref{53}) with the help of the Poisson formula, we find
\begin{equation}
\rho(E)={1\over 2\pi l_B^ 2}\sum_{N=0}^{\infty}
\sqrt{{m\over\Gamma E}}\exp\left\{-{\pi m\over\Gamma E}
[E- \omega_c(N +{1/2})]^2\right\}
\label{55}
\end{equation}
As in eq.(\ref{37}), the Landau levels have a Gaussian shape. However,
in contrast to the case of random potential, their width does not
increase with the magnetic field, but is instead proportional to
$E^{1/2}$. When the oscillations are relatively weak, they can be
characterized by the first harmonic with an amplitude
\begin{equation}
\rho_{osc}\simeq -{m\over\pi}\cos\left({2\pi E\over\omega_c}\right)
e^{-\pi E\Gamma/m\omega_c^2}
\label{56}
\end{equation}

These results for the amplitude of oscillations can be generalized to
the SdHO of conductivity, as in the preceding section. We get again
eq.(\ref{41}), with $S_W$ replaced by $S_h$, eq.(\ref{54}). We can
also consider a random magnetic field with finite correlation length
$\xi$ and correlator
\begin{equation}
\langle h(\bbox{r})h(\bbox{r'})\rangle = U(|\bbox{r}-\bbox{r'}|)
\label{57}
\end{equation}
We find then
\begin{eqnarray}
\rho_{osc},\ \sigma_{osc}&\propto& e^{-S_h}\ ; \nonumber\\
S_h&=&{1\over 2}\int_{|\bbox{r}|\le R_c} d^2r
\int_{|\bbox{r'}|\le R_c} d^2r' U(|\bbox{r}-\bbox{r'}|)=
\pi R_c^2\int_0^\infty {dq\over q} \tilde{U}(q) J_1^2(qR_c) \nonumber\\
&\simeq& \left\{
\begin{array}{ll}
\displaystyle{\frac{\pi E \int d^2 r U(r)}{m\omega_c^2}}\ ,
\qquad & R_c\gg\xi\\
\displaystyle{\frac{2\pi^2E^2U(0)}{m^2\omega_c^4}}\ ,\qquad &  R_c\ll\xi\ ,
\end{array}
\right.
\label{58}
\end{eqnarray}
where $\tilde{U}(q)$ is the Fourier transform of $U(r)$.
Therefore, the Dingle plot will be quadratic for $R_c\gg\xi$ and
quartic in the opposite case.

\section{Magnetooscillations near the $\nu=1/2$ filling factor of the
Landau level.}

In this section, we consider a realistic model describing the electron
gas in $GaAs--AlGaAs$ heterostructures where the FQHE is observed. The
system is formed by the 2D electron gas of density $n_e$ and by the
positively charged impurities located in a layer separated by a large
distance $d_s$ from the electron plane. Each impurity creates a
potential of the form
\begin{equation}
\int(dq)v_{0}(q)e^{i\mbox{\boldmath$q$}(\mbox{\boldmath$r$}-
\mbox{\boldmath$r$}_{i})}\ ; \qquad
v_{0}(q)={2\pi e^{2}\over \epsilon q}e^{-qd_{s}},
\label{59}
\end{equation}
where $\bbox{r}_{i}$ is the projection of the impurity position
to the 2D plane and  $\epsilon$ is the dielectric constant.
We briefly consider first the magnetooscillations in this system in
low magnetic fields. The charge carriers are then characterized by
their lattice mass $m_b$, which for $GaAs$ is equal to $m_b\simeq 0.07
m_e$, $m_e$ being the free electron mass. The potential $v_0(q)$,
eq.(\ref{59}), gets screened by the 2D electron gas into
\begin{equation}
v(q)={2\pi \over m_b}e^{-qd_{s}}.
\label{60}
\end{equation}
When writing eq.(\ref{60}), we assumed that the DOS determining the
screening is practically constant: $\rho\approx m_b/2\pi$. Taking into
account the magnetooscillations of the DOS here would lead to a non-linear
screening, see \cite{efros}. However, since
our primary interest is in the region of large $p$ where the amplitude
of oscillations is small, we can neglect this non-linear effect.
We also assume the impurity positions to be uncorrelated (see,
however, the discussion in the next section). Then we find that the total
random potential of the impurities is described by the correlator
$W(r)$, see eq.(\ref{1}), with the Fourier transform
\begin{equation}
\tilde{W}(q)=n_i\left({2\pi\over m_b}\right)^2e^{-2qd_s}
\label{61}
\end{equation}
According to section 3, we find then for the amplitude of oscillations
\begin{equation}
\rho_{osc},\ \sigma_{xx}^{osc}\propto - \cos\left({2\pi^2n_ec\over
eB}\right)e^{-S_W} \ ,
\label{62}
\end{equation}
with $S_W$ given by eq.(\ref{34}). In particular, for $R_c\gg d_s$ we
find
\begin{eqnarray}
&&\rho_{osc},\ \sigma_{xx}^{osc}\propto - \cos\left({2\pi^2n_ec\over
eB}\right)\exp(-\pi/\omega_c\tau_0) \ ; \nonumber\\
&&{1\over\tau_0}={1\over\pi v_F}\int dq\tilde{W}(q)=
{n_i\over m_b d_s}\left({2\pi\over n_e}\right)^{1/2}\ ,
\label{63}
\end{eqnarray}
whereas for $ R_c\ll d_s$
\begin{eqnarray}
&&\rho_{osc},\ \sigma_{xx}^{osc}\propto - \cos\left({2\pi^2n_ec\over
eB}\right)\exp\left[-\left(\pi/\omega_c\tau^*_0\right)^2\right] \ ;
\nonumber\\
&&{1\over\tau^*_0}=[2W(0)]^{1/2}=
{1\over m_b d_s}({\pi n_i})^{1/2}\ .
\label{64}
\end{eqnarray}
For the systems under consideration the usual assumption is that
concentrations of donors and charge carriers coincide: $n_e=n_i$. Then
the condition of weak oscillations $\omega_c\tau_0\ll 1$ reduces to
$R_c\gg d_s$. Therefore, in the region of small oscillations (where
our derivation is justified), eq.(\ref{63}) has to be applied.

Now we consider the oscillations in strong magnetic field, near half
filling of the Landau level: $\nu=2\pi c n_e/eB\simeq 1/2$. It was
observed experimentally that in this region the longitudinal
resisitivity shows oscillations with minima at $\nu=p/(2p\pm 1)$,
very much reminiscent to its behavior in low magnetic fields where
conventional SdHO take place. This feature did not find an explanation
within the original hierarchy theory of the FQHE \cite{laugh,hal1}.
To explain it, Jain
\cite{jain} proposed a concept based on converting the electrons into
composite fermions by attaching to them an even number of flux quanta.
A field-theoretical formalism based on  the Jain's idea was developed
by Lopez and Fradkin \cite{lofra}. In this approach, the statistical
transformation of electrons into composite
fermions is implemented by introducing a Chern--Simons (CS) gauge
field interacting with electrons.

Following a similar approach Halperin, Lee and Read \cite{hlr}
developed a theory for the half filled Landau level (see also
\cite{kz}).
This theory gives an explanation for many experimentally
observed properties of the $\nu=1/2$ state,
such as a non-zero value of the longitudinal resistivity,
an anomaly in the surface acoustic wave propagation \cite{will}, and
a dimensional resonance of the composite fermions \cite{kan}.
It predicts the formation, at half filling, of a metallic state with
well defined Fermi surface. From this point of view,
the $\nu=p/(2p\pm 1)$ series can be considered
as the usual  $\nu=p$ Shubnikov--de Haas oscillations (SDHO) for the
composite fermions, providing an explanation for the prominence of the
 above FQHE states.

 Details of the CS gauge field formalism can be found
e.g. in \cite{lofra,hlr} and are not presented here.
The statistical transformation attaches to each electron
an even number $\tilde{\phi}$ of flux quanta of the CS gauge field.
To describe the vicinity of the  $\nu=1/2$ state, we take
$\tilde{\phi}=2$; the same formalism with
$\tilde{\phi}=4$ can be applied to the $\nu=1/4$ state.
In the mean field approximation, the statistical magnetic field
$B\sb{1/2}=4\pi c n\sb{e}/e$ cancels exactly the externally
applied field $B$ at $\nu=1/2$.
When the filling factor $\nu$ is tuned away from $\nu=1/2$,
the effective uniform magnetic field is equal to
$B\sb{e\!f\!f}=B-B\sb{1/2}$.
For  $\nu$ close to $1/2$, the number of filled Landau levels of
composite fermions
$p \gg 1$, so that the problem can be considered quasiclassically.

Although a static impurity creates a scalar potential (\ref{59}) only,
it acquires also a vector component due to screening by fermions and
mixing with the CS field. In the random phase arrroximation one gets
\begin{equation}
a\sb{\mu}=\left(\delta\sb{\mu}\sp{\
\rho}-U\sb{\mu\nu}K\sp{\nu\rho}\right)\sp{-1}
a\sb{\rho}\sp{(0)},
\label{dais}
\end{equation}
where we united scalar $a\sb{0}$ and vector $\bbox{a}$ potentials
in a covariant vector $a\sb{\mu}$; the vector $a\sb{\rho}\sp{(0)} $
represents the bare impurity potential, eq.(\ref{59}) \
and therefore has only $\rho=0$ non-zero component.
The tensors $U\sb{\mu\nu} $ and $K\sp{\nu\rho}$ represent the
bare gauge field propagator and the current-density response tensor
of the composite fermions, respectively.

To evaluate eq.(\ref{dais}) we use the Coulomb gauge $\mbox{div}\bbox{a}=0$,
go to the momentum space and choose the momentum $\bbox{q}$ to be
directed along the $x$-axis: $q\sb{x}=q$, $q\sb{y}=0$.
Then  $a\sb{\mu}$ has only 2 non-zero components corresponding to
$\mu=0$, $y$, and both K and U become $2\times 2$ matrices
\cite{hlr}:

\begin{eqnarray}
&&K\sp{\mu\nu}(q)=
\left(\begin{array}{cc}
-m\sp{*}/2\pi & -iq\sigma\sb{xy}\\ iq\sigma\sb{xy} & \chi
q\sp{2}-2i\omega n\sb{e}/qk\sb{F}
\end{array}\right) \nonumber \\
&&U\sb{\mu\nu}(q)=
\left(\begin{array}{cc}
v(q)  & 2\pi i \tilde{\phi}/q\\-2\pi i \tilde{\phi}/q  & 0
\end{array}\right) \label{tens} \\
&&a\sb{\mu}\sp{(0)}(q)=
\left(\begin{array}{c}
 v\sb{0}(q)e\sp{-i\mbox{\boldmath$qr$}\sb{i}}  \\ 0
\end{array}\right) \ , \nonumber
\end{eqnarray}
where $m\sp{*}$ is the effective mass of fermions,
$\chi=1/24\pi m\sp{*}$ is the magnetic susceptibility,
$v(q)=2\pi e\sp{2}/(\epsilon q)$  is the Coulomb propagator,
and $\sigma\sb{xy}$ is the Hall conductivity of composite fermions.

Substituting (\ref{tens}) in (\ref{dais}), we find
\begin{equation}
a\sb{\mu}(q)=
{v\sb{0}(q)e\sp{-i\mbox{\boldmath$qr$}\sb{i}}\over
{ m\sp{*}v(q)\over 2\pi} + (\tilde{\phi}s+1)\sp{2}
+{\tilde{\phi}\sp{2}\over 12}}
\left(\begin{array}{c}
\tilde{\phi}s+1  \\ i\tilde{\phi} m\sp{*}/q
\end{array}\right),
\label{scr}
\end{equation}
where $s=2\pi\sigma\sb{xy}\simeq p$ in the limit
$\omega\sb{c}\tau\sb{t}\gg 1$ .

Depending on relations between parameters of the problem, one can find
various regimes of behavior of the oscillations amplitude. We will
concentrate on a regime which is the most relevant to the experiment.
Let us  compare the first and the second term in denominator of
(\ref{scr}).
As we will see below, the typical momenta are $q\sim (2d\sb{s})\sp{-1}$,
and we get for $\tilde{\phi}=2$
\begin{equation}
{ m\sp{*}v(q)/2\pi\over (2s)\sp{2}}=
{m\sp{*}e\sp{2}\over 4\epsilon q s\sp{2}}\sim
{m\sp{*}e\sp{2}\over\epsilon k\sb{F}}{k\sb{F}d\sb{s}\over2p\sp{2}}\sim
{50\over p\sp{2}},
\label{est}
\end{equation}
where $k\sb{F}=\sqrt{4\pi n\sb{e}}$,
and we used  typical experimental parameters
\cite{st2} $n\sb{e}=1.1\cdot 10\sp{11}\mbox{cm}\sp{-2}$,
$d\sb{s}=80nm$, and the experimentally estimated value for the ratio
${m\sp{*}e\sp{2}/(\epsilon k\sb{F})}\sim 10$.
For the not too large $p$ we are interested in,
it is thus a reasonable approximation to neglect all but the first
term in the denominator of (\ref{scr}). This gives
\begin{equation}
a\sb{\mu}(q)=
{2\pi\over
 m\sp{*}} \tilde{\phi}e\sp{-i\mbox{\boldmath$qr$}\sb{i}}e\sp{-qd\sb{s}}
\left(\begin{array}{c}
p  \\ i m\sp{*}/q
\end{array}\right),
\label{scr1}
\end{equation}

The random field action $S\sb{r}$ (analogous to $S_W$ in section 3 or
$S_h$ in section 4)
is given by
\begin{equation}
S_{r}={1\over 2}\left\langle\left(\oint{a\sb{\mu}dr\sp{\mu}}\right)^2
\right\rangle=
{1\over 2}\left\langle\left(\int a\sb{0}dt -\oint
\bbox{a}d\bbox{r}\right)^2
\right\rangle\ ,
\label{72}
\end{equation}
where the integration goes around a cyclotron orbit.
Averaging over the impurity configurations, we find
\begin{eqnarray}
&&S\sb{r}= (2\pi\tilde{\phi})\sp{2}n\sb{i}
\nonumber \\
&&
\times\int (dq)e\sp{-2qd\sb{s}}
\left|{p\over k\sb{F}}\oint dl\:
e\sp{-i\mbox{\boldmath$qr$}}+
\int d\sp{2}r\: e\sp{-i\mbox{\boldmath$qr$}} \right|\sp{2}
\label{act}  \\
&&=n\sb{i}(4\pi\sp{2}\tilde{\phi} R\sb{c})\sp{2}
\int (dq)e\sp{-2qd\sb{s}}
\left|{p\over k\sb{F}}J\sb{0}(qR\sb{c})+
{1\over q}J\sb{1}(qR\sb{c})\right|\sp{2}
\nonumber
\end{eqnarray}
Here $\oint dl$ means integration along the cyclotron orbit
and corresponds to the electric field contribution,
whereas $\int d\sp{2}r$
goes over the area surrounded by the orbit and describes the magnetic
field contribution.
Taking into account that $R\sb{c}\sp{2}=p\sp{2}/(\pi n\sb{e})$,
we have $R\sb{c}/2d\sb{s}=p/\sqrt{4\pi n\sb{e}d\sb{s}\sp{2}}\sim p/10
\lesssim 1$.
Thus for relevant momenta $q\sim1/(2d\sb{s})$ and  level numbers $p$,
 $qR\sb{c}\ll 1$ is a reasonable approximation.
In this case eq.(\ref{act}) reduces to
\begin{equation}
S\sb{r}=
\pi\sp{3}\tilde{\phi}\sp{2}n\sb{i}
{R\sb{c}\sp{4}\over d\sb{s}\sp{2}}=
{n\sb{i}\over n\sb{e}}
{\pi\tilde{\phi}\sp{2}\over n\sb{e}d\sb{s}\sp{2}}
\left( {2\pi n\sb{e}\over m\sp{*}\omega\sb{c}}\right)\sp{4}.
\label{act2}
\end{equation}
Note that electric and magnetic field fluctuations give equal
contributions in this limit.
According to sections 3,4, this gives for the oscillating part
of the conductivity:
\begin{equation}
\sigma\sb{xx}\sp{osc}\propto -
\cos{\left({4\pi\sp{2}n\sb{e}c\over e B\sb{e\!f\!f}}\right)}
\exp{
\left[-\left({\pi\over\omega\sb{c}\tau\sb{1/2}\sp{*}}\right)\sp{4}\right]}
\label{sxxa}
\end{equation}
where we introduced a parameter $\tau\sb{1/2}\sp{*}$
which is given according to eq.(\ref{act2}) by
\begin{equation}
\tau\sb{1/2}\sp{*}\simeq { m\sp{*}\over 2}
\left({d\sb{s}\sp{2}\over 4\pi n_i n_e^2}\right)
\sp{1/4}\ .
\label{tau}
\end{equation}

Let us briefly consider now effect of finite temperature $T$. First of
all, the SdHO are then suppressed by the usual factor
$D_T=(2\pi\sp{2}T/\omega\sb{c})/\sinh(2\pi\sp{2}T/\omega\sb{c})$
originating from the Fermi distribution \cite{abrikos}. In addition, the
fermions are scattered by the thermal fluctuations of the gauge
field.
The propagator of gauge field fluctuations is given by
\begin{equation}
D\sb{\mu\nu}(q,\omega)=U\sb{\mu\rho}(q)
\left(\delta\sp{\rho}\sb{\ \nu}-
K\sp{\rho\lambda}(q,\omega)U\sb{\lambda\nu}(q)\right)\sp{-1}
\label{d}
\end{equation}
In particular for the $D\sb{11}$ component determining the magnetic
field fluctuations, we get
\begin{eqnarray}
&&D\sb{11}(q,\omega)=
(2i\omega n\sb{e}/ q k\sb{F}-\tilde{\chi}q\sp{2})\sp{-1} ;
 \nonumber \\
&&\tilde{\chi}=
{1\over2\pi m\sp{*}}\left[{1\over
12}+\left(s+{1\over\tilde{\phi}}\right)\sp{2}
\right]+{v(q)\over (2\pi\tilde{\phi})\sp{2}}.
\label{d11}
\end{eqnarray}
In the quasistatic approximation we find
\begin{equation}
\langle A\sb{1}A\sb{1}\rangle\sb{q}=
\int{d\omega\over 2\pi}{2T\over\omega}\mbox{Im}D\sb{11}=
{T\over\tilde{\chi} q\sp{2}}
\label{79}
\end{equation}
and consequently for the amplitude of
magnetic field fluctuations
\begin{equation}
\left< hh\right>\sb{q}=T/\tilde{\chi}.
\label{80}
\end{equation}
If $\omega_c\tau_t\gg 1$, we have $s\simeq p$ and
$\tilde{\chi}=12p^2\chi$. Therefore, the effective magnetic
susceptibility $\tilde{\chi}$ much exceeds its bare value $\chi$,
leading to a strong suppression of the fluctuations (\ref{79}),
(\ref{80}). A similar suppression of the gauge field fluctuations in
external magnetic field was found in \cite{iw} where the
magnetoconductivity of the doped Mott insulators was studied.
The contribution of the fluctuations (\ref{80}) to the random field
action $S_r$, eq.(\ref{72}), is
\begin{equation}
 S_r^{(T)}={T\over\tilde{\chi}}\pi R_c^2\approx
{2\pi^2\over p} {T\over\omega_c}\ ,
\label{81}
\end{equation}
i.e. is small at $p\gg 1$ compared to the standard term $-\ln D_T$  and
decreases with $p$.

The slope of the dependence of $\ln\rho_{osc}$ on $1/B$ at
$T\gg\omega_c$ is conventionally used to extract a value of the
effective mass $m^*$ \cite{shoenberg}.
This procedure is based on the assumption that
$\rho_{osc}\propto D_T$, so that $\ln\rho_{osc}=2\pi^2Tm^*c/eB +
const$. The additional attenuation factor $\exp(-S_r^{(T)})$,
eq.(\ref{81}), would then lead to a fictitious $1/p$ correction to the
effective mass resulting in its apparent decrease with $p$ at
moderately large $p$.

Let us compare our findings with available experimental results.
In Fig.1 we present low--temperature
experimental data for the amplitude
of $\rho\sp{osc}$ from \cite{st2}
($T=0.19K$, $B\sb{e\!f\!f}>0$).
It is seen that they can be fitted well by
$\exp\left[-(\pi/\omega\sb{c}\tau^*)\sp{4}\right]$,
whereas a conventional
$\exp(-\pi/\omega\sb{c}\tau)$ fit is much worse.
Therefore the data apparently show the behavior
$\ln\rho_{osc}\propto 1/\omega_c^4$ predicted by eq.(\ref{sxxa}). The
value of the parameter $\tau_{1/2}^*$ which is found from such a fit
is $\tau_{1/2}^*=16\cdot 10^{-12}s$
At the same time the theoretical estimate according to eq.(\ref{tau})
(with use of the parameters of \cite{st2}) gives
$\tau\sb{1/2}\sp{*}\simeq 2.4\cdot 10\sp{-12}s$
if one uses the experimental value of
the effective mass $m\sp{*}=0.7m\sb{e}$.
A similar discrepancy is found for the low--field relaxation time:
eq.(\ref{63}) for $m\sb{b}=0.07m\sb{e}$ gives
$\tau\sb{0}\sp{*}\simeq 0.6\cdot 10\sp{-12}s$,
whereas the value quoted in \cite{st2}
is $\tau\sb{0}\simeq 9\cdot 10\sp{-12} s$.
We note also that the theoretically estimated values for the transport
relaxation rate at $\nu=1/2$ are typically $4$ times greater
than extracted from  experimental mobilities \cite{hlr,st2}.
Therefore the theory seems to overestimate relaxation
rates systematically.
 This situation has been discussed previously
\cite{col,lev}. The considerable increase of relaxation times was
attributed to the correlations in positions of charged impurities
due to their mutual Coulomb interaction \cite{lev}.
We still encounter a problem, however, when trying to explain the
above discrepancy in the value of $\tau_{1/2}^*$ in this way. The
correlations between impurities lead to a damping of the correlator
$\tilde{W}(q)$, eq.(\ref{61}) by a certain $q$--dependent factor.
Since the mechanism of suppression of SdHO and the characteristic
momenta are the same near $B=0$ and $\nu=1/2$, we expect a suppression
of the action $S_W$ in (\ref{62}) and $S_r$ in (\ref{act}) by roughly
the same factor. For the case of low fields this factor is
$\tau_0(exp.)/\tau_0(theor.)\simeq 15$. At the same time, to reconcile
the experimental data in the vicinity of $\nu=1/2$ with
eq.(\ref{tau}), we need this factor to be
$[\tau_{1/2}^*(exp.)/\tau_{1/2}^*(theor.)]^4\simeq 2300$, i.e 150
times larger! It is not clear to us, what could be the source of such a
drastic weakening of the random fields.

More recently, experimental data \cite{st3} on SdHO near $\nu=1/2$ on
a better quality sample have been published. The obtained Dingle plot
[Fig.3(a) of Ref.\cite{st3}] is again highly nonlinear and can be
well fitted by our formula (\ref{sxxa}). The fit yields in this case
the value $\tau_{1/2}^*=9.5\cdot 10^{-12}s$, whereas the theoretical
estimate according to eq.(\ref{tau}) using the appropriate values of
$n_e$,$d_s$ and $m^*$ gives $\tau_{1/2}^*=2.0\cdot 10^{-12}s$. The
discrepancy is somewhat smaller than for the sample from \cite{st2}
but still very large.

Now we discuss experimental data at higher temperatures. As a result
of their analysis, an experimental value of the effective mass as a
function of $p$ was obtained in \cite{st2,le,st3}. At moderately large
$p$ $m^*$ was found to be slowly decreasing with $p$ in agreement with
our results. For larger $p$, a sharp increase of $m^*$ was observed in
\cite{st3}, the origin of which is not clear to us. Let us note that the
suggestion made in \cite{st3} to explain the non-linearity of the
Dingle plot  by this variation of the effective
mass is not supported by our results, since $m^*$ drops out from
eq.(\ref{sxxa}).

\section{Conclusions.}

We have shown in this article that the path integral formalism
in the quasiclassical approximation allows
one to study the density of states and the conductivity of a system
of charge carriers scattered by a long-range correlated random
potential or a random magnetic field. We focussed attention on the
oscillatory behavior of these quantities as a function of an applied
magnetic field. For a random potential of correlation length $\xi$, we
found the magnetooscillations in both quantities to be attenuated
exponentially, the exponent being $\propto \omega_c^{-1}$ for
$R_c\gg\xi$ and $\propto \omega_c^{-2}$ for
$R_c\ll\xi$, where $\omega_c$ is the cyclotron frequency and $R_c$ the
cyclotron radius in the external magnetic field. In the case of a
random magnetic field of correlation length $\xi_B$, the
magnetooscillations are again exponentially damped, this time with exponent
$\propto \omega_c^{-2}$ for
$R_c\gg\xi_B$ and $\propto \omega_c^{-4}$ for
$R_c\ll\xi_B$. Our results show that the amplitude of
magnetooscillations as a function of the magnetic field can be used to
identify the scattering potential or random magnetic field.

Finally, we considered the FQHE system near $\nu=1/2$ in the composite
fermion picture. Here the Chern--Simons field of the flux tubes
attached to every fermion gives rise to static random field
fluctuations at impurity sites, where fermions may be trapped.  For
experimentally relevant values of input parameters, the damping of
magnetooscillations is found to be proportional to
$\exp[-(\pi/\omega_c\tau_{1/2}^*)^4]$.
This is in good agreement with experimentally obtained
highly non-linear Dingle plots which can be very well fitted by a
quartic dependence. However, the experimental value of the parameter
$\tau_{1/2}^*$ is about 8 times greater than our theoretical
estimate. This means that the random fields are much weaker than we
expect them to be. Taking into account correlations between impurities
seems not to resolve this discrepancy, as we discussed in the end of
the preceding section. It is not clear to us at present whether this
additional weakening of random fields can be explained within the
composite fermions theory or else implies a certain inconsistency of
this theory.

This work was initiated by the late Arkady Aronov, and was partially
done in collaboration with him. We very much regret that he did not
live to see the completion of the work. We are grateful to Yehoshua
Levinson for useful discussions.
This work was supported by the Alexander von Humboldt Stiftung
(A.D.M.), SFB 195 der Deutschen
Forschungsgemeinschaft (A.D.M. and P.W.)
and by the German-Israel Foundation for Research (E.A.)

\begin{figure}
\caption{Dingle plot. Logarithm of normalized  amplitude of resistivity
oscillations $\ln({D\sb{T}\rho\sp{osc}/4\rho})$, with
$D\sb{T}=\sinh(2\pi\sp{2}T/\omega\sb{c})/(2\pi\sp{2}T/\omega\sb{c})$,
as a function of inverse effective magnetic field $B\sb{e\!f\!f}\sp{-1}$.
Experimental data from \protect\cite{st2} (squares) and
\protect\cite{st3} (circles) are presented, as well as their fits with
eq.(\protect\ref{sxxa}).}
\end{figure}
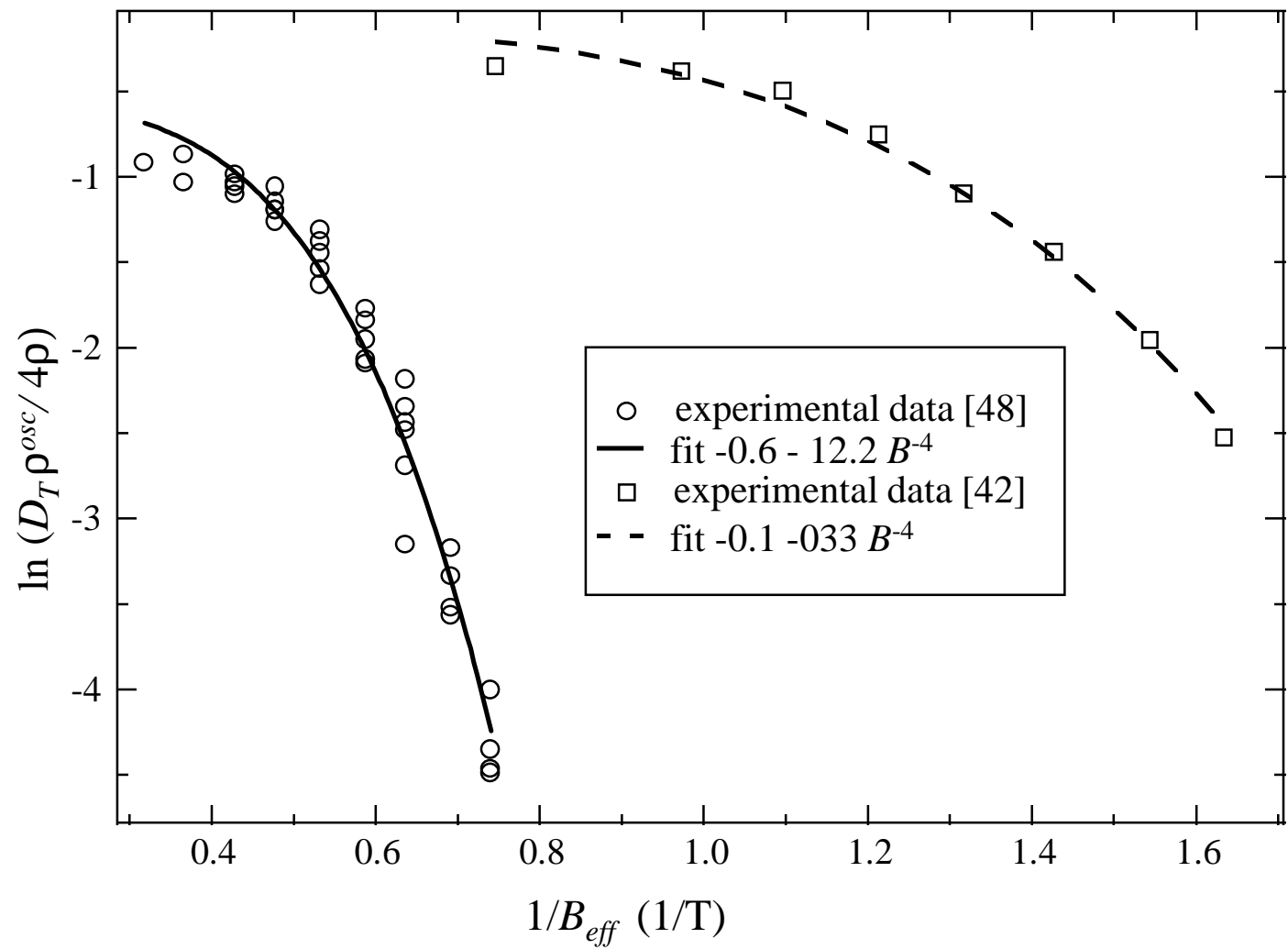
\end{document}